\documentclass[reprint, amsmath,amssymb, showpacs, aps]{revtex4-1}

\usepackage{graphicx}% Include figure files
\usepackage{dcolumn}% Align table columns on decimal point
\usepackage{bm}% bold math
\begin{document}

\preprint{140407CCDcal.tex}

\title{Absolute calibration of a CCD camera with twin beams}


\author{I. Ruo-Berchera}
 \email{i.ruoberchera@inrim.it}
\author{A. Meda}%
\author{I. P. Degiovanni}%
\author{G. Brida}%
\author{M. L. Rastello}
\author{M. Genovese}%

\affiliation{Istituto Nazionale di Ricerca Metrologica, Strada delle Cacce 91, 10135 Torino, Italy}


%\affiliation{
%\affiliation{
%\affiliation{%


\date{\today}% It is always \today, today,

\begin{abstract}
 We report on the absolute calibration of a CCD camera by exploiting quantum correlation. This novel method exploits a certain number of spatial pairwise quantum correlated modes produced by spontaneous parametric-down-conversion. We develop a measurement model %taking into account all the possible source of losses and noise that are not related to the quantum efficiency,
accounting for all the uncertainty contributions, and we reach the relative uncertainty of 0.3$\%$ in low photon flux regime. This represents a significant step forward for the characterizaion of (scientific) CCDs used in mesoscopic light regime. %(i.e. ?????$\div$????  photons (s $ \times$ pixel) $^{-1}$ ).
\end{abstract}

\pacs{Valid PACS appear here}% PACS, the Physics and Astronomy
\maketitle

\emph{Introduction.}
The development of quantum metrology, imaging and sensing \cite{qm1,qm2,qm3, qm4, qm5,BrambPRA2004} based on quantum optical states aim to reach sensitivity beyond classical limits. For this reason it requires new detection strategies in the low light intensity regime (down to few photons), that are far from the ones where traditional radiometry operates \cite{fox}. Development of dedicated methods for absolute calibration of detectors in this context is therefore necessary, as it is widely recognised inside the radiometric community \cite{qcan1}. Some specific activities in this context are already on-going\cite{qcan2,MIQC}, in particular related to the calibration of single-photon detectors exploiting the Klyshko's twin-photon technique \cite{bp1,b,a,b1,Brida2000a, PolMig2007, Cheung2011,Polyakov2009} and its developments \cite{n, pe,aga,s,v, Worsley2009, Schmunk2011}.

In particular, demand for precise calibration of detectors both in mesoscopic light regime it is relevant not only from the point of view of the development of quantum technologies, but also for establishing a connection between the light intensity level typical of classical radiometric measurements and the quantum radiometry operating at single-photon level \cite{qcan1,qcan2}. Quantum correlations in twin beams offer an opportunity for reaching this goal, as discussed in  \cite{BridaJOSAB2006, BridaOE2008, BridaJMO2009, RuoASL2009}.

In \cite{Brida2010} we realized the first absolute calibration of a standard CCD camera by exploiting bright squeezed vacuum. Standard CCD cameras, i.e. without any avalanche electro-multiplication, are able to count the number of generated photo-electron in each pixels for a given exposure time with a read-out noise $\Delta_{RN}$ of few photo-electron (for top-level devices), independent of the exposure time. However, a single photon can not be distinguished from the noise and the time resolution does not allow time tagging and coincidence of photon arrivals, instead the signal is proportional to the intensity (analog regime). If a light signal generate N photo-electron, under the condition  $\Delta_{RN}<<N^{1/2}$, sub-shot-noise sensitivity can be achieved \cite{JedrkPRL2004,BridaPRL2009,BridaNP2010}, a properties that has been used also for quantum enhanced sensing and imaging protocols \cite{qm2, BridaNP2010}.

The method in \cite{Brida2010} is based on the sub-shot-noise measurement of photon number correlation between a pair of spatio-temporal multimode twin beams. According to quantum mechanics, perfect correlation are generated in twin beams, however losses on the optical path represent a random process that lower in a calculable way the average correlation. Therefore, quantum efficiency can be estimated by the the actual measured correlation level. However, the uncertainty reported in that work was still too large to demonstrate the validity of the method for metrological purposes. In this paper we report a refined method, for calibrating a standard CCD, improving the uncertainty of order of magnitude with respect to \cite{Brida2010}, which is aligned with the state of the art of the absolute calibration of single photon detector with Klyshko's method \cite{PolMig2007, Cheung2011}, allowing the same optical source and experimental setup to be used for calibrating and comparing detector in a completely different regime. Together with an overall improvement of the stability and reduction of noise sources in the experiment, the main improvement of the present realization is represented by the theoretical model, which allows decoupling the losses contribution due to misalignment and collection of correlated modes with the one due to the quantum efficiency, thus allowing to introduce the appropriate correction. A precise estimation of the coherence radius of the modes and an accurate positioning of the pixel array with respect to the physical center of symmetry of the twin beam represent the key elements of the present improvement of the absolute calibration technique.

\emph{Experimental set up.} The setup is composed by a 406nm laser operating in quasi-CW mode, i.e. with pulse duration of 70 ms synchronized with the exposure time of the CCD camera (a Princeton Inst. Pixis 400BR), at a power of 15 mW. The laser is spatially filtered in order to remove non Gaussian components, collimated, and then addressed to a $5\times5\times15$ mm$^3$ BBO non-linear crystal of Type II, where the two correlated beams are produced.

After the crystal two mirrors reflect the generated twin beam (TWB) towards the CCD camera, while the pump beam passes between the mirrors (this solution presents the benefit of avoiding the creation of noisy fluorescence light by pump-stopping filters). A far field lens with a focal length of $f=10$cm is located  in a $f-f$ configuration between the output surface of the crystal and the detection plane. It is followed by a couple of polarization plates so that the two correlated beams with orthogonal polarization are isolated each other at the camera. By a 10 nm FWHM interference filter we chose a proper phase matching configuration similar to the one  reported in Fig. \ref{setup}(a) (right-hand-side), which allows maximizing the area around degenerate frequency ($\lambda=812$ nm) available for the measurement. Then we remove the narrow filter, obtaining the typical single shot image presented in Fig. \ref{setup}(b). All the analysis has been performed adopting a hardware binning of the physical pixels of the CCD in groups of $24 \times 24$. The resulting super-pixels have  a size $l_{pix}=480$ $\mu$m.

\begin{figure}[tbp]
\begin{center}\label{setup}
\includegraphics[angle=0, width=8cm]{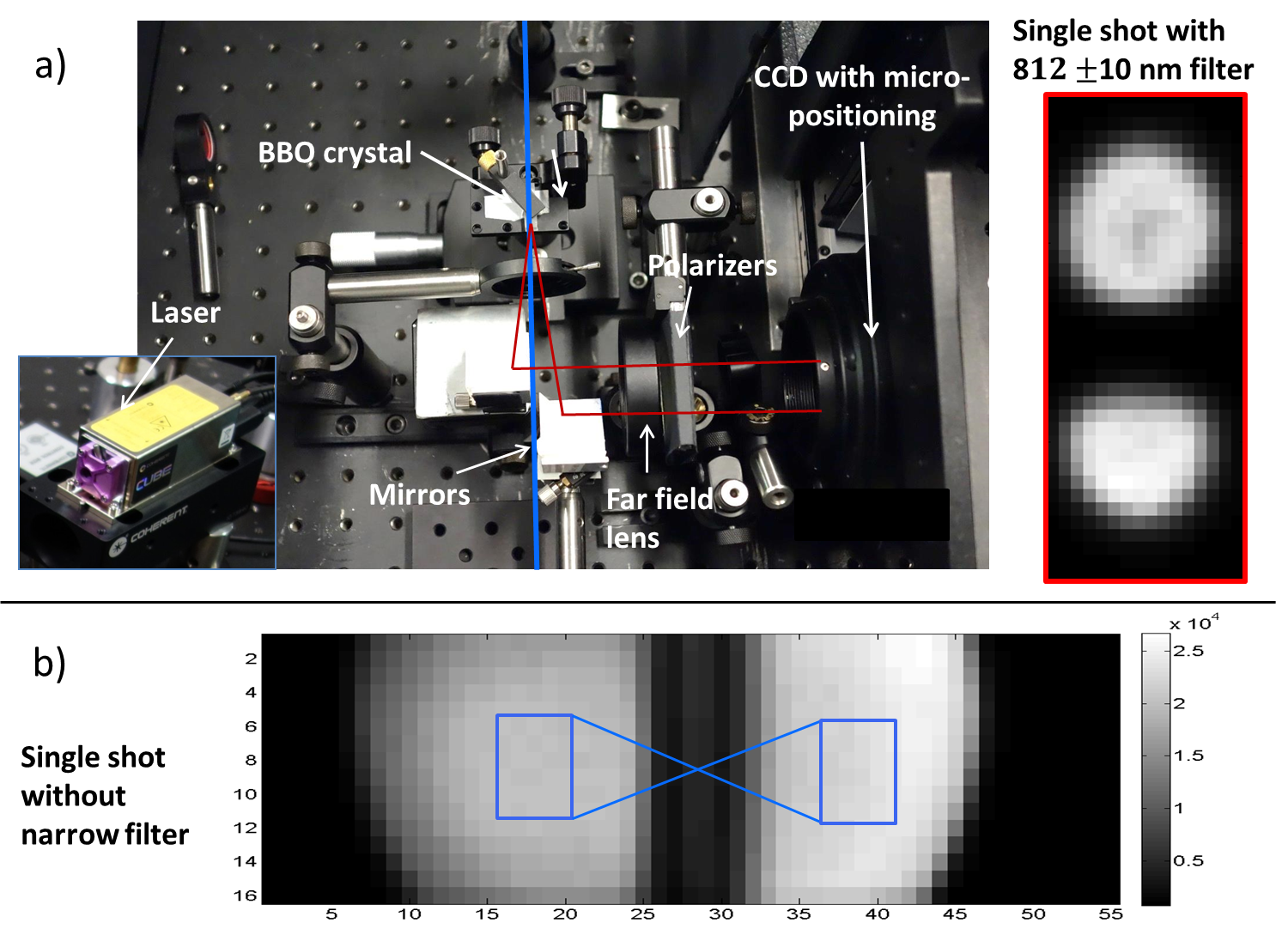}
\caption{a) Final part of the experimental setup. The box on the right shows the phase-matching configuration corresponding to the 812 nm degenerate wavelength. b) Image of the TWB produced by a single laser shot without narrow filtering. Only a long pass filter, $\lambda >750$ nm, is used. The number of photons detected is reported in the gray scale bar on the right. \label{setup}}
\end{center}
\end{figure}

\emph{Theoretical model.} The state produced by spontaneous PDC in non-linear crystals, in the approximation of plane
wave pump field of frequency $\omega_{p}$ and propagating in the $z$
direction, can be expressed as a tensor product $|\Psi\rangle=\bigotimes_{\mathbf{q},\Omega}|\psi(\mathbf{q},\Omega)\rangle$ of two-mode photon-number-entangled states of the form
\begin{equation}\label{two-mode state}
|\psi(\mathbf{q,\Omega })\rangle=\sum_{n}C_{\mathbf{q,\Omega}}(n)|n\rangle_{\mathbf{q,\Omega}}|n\rangle_{-\mathbf{q},-\Omega},
\end{equation}
where the modes identified by the frequencies $\omega_{p}/2\pm\Omega$ and transverse
momenta $\pm\mathbf{q}$, are entangled in the photon number basis $\left\{|n\rangle\right\}$. In the Type II Crystals in particular, the two modes $({\mathbf{q},\Omega})$ and $({-\mathbf{q},-\Omega})$ are also distinguished by orthogonal polarization. Hereinafter we will label, as usual, the two correlated modes as signal ``$s$" and idler ``$i$".

The coefficient $C_{\mathbf{q,\Omega}}(n)$ is related to the mean number of photon in the mode $\langle n_{\mathbf{q,\Omega}}\rangle$,  which can be considered constant over a broad spatio-temporal bandwidth, i.e. $C(n) \propto \sqrt{\mu^{n}/(\mu+1)^{n+1}}$, where  $\langle n_{\mathbf{q,\Omega}}\rangle = \mu$.

In the far field region (obtained as the focal plane of a thin lens of
focal length $f$ in a $f-f$ configuration, see Fig. 1), any
transverse mode $\mathbf{q}$ is associated with a single position
$\mathbf{x}$ in the detection (focal) plane according to the
geometric transformation $(2c f/\omega_{p})\mathbf{q}\rightarrow
\mathbf{x}$, with $c$ the speed of light. Therefore, a perfect
correlation appears in the photon number $n_{i,\mathbf{x}}$ and
$n_{s,-\mathbf{x}}$ registered at the degeneracy frequency $\Omega=0$ by two detectors placed in two
symmetrical positions $\mathbf{x}$ and $-\mathbf{x}$, where the
center of symmetry (CS) is basically the pump-detection plane
interception (assuming negligible walk-off).

In real experiments the pump laser is not a plane wave, rather it
can be reasonably represented by a gaussian distribution with
spatial waist $w_{p}$, inducing an uncertainty in the relative
propagation directions of the twin photons of the order of the
angular bandwidth of the pump,  $[(2\pi c
f)/(\omega_{p}w_{p})]^{2}$ \cite{BrambPRA2008}. Moreover, the non-null frequency bandwidth (about 20 nm in our experiment) determines a further broadening of the spot in which correlated detection events occur. These effects contribute to the actual size of the  cross-correlation (or coherence) area corresponding to the mode's size in the far field.
In order to collect all correlated photons, the detection area $\mathcal{A}_{det}$ must be
larger than the coherence area $\mathcal{A}_{coh}$, \cite{BrambPRA2008}).

The ideal situation of detecting photons over two equal and
symmetric areas, $\mathcal{A}_{det,j}$ ($j=s,i$), containing a large number of transverse spatial modes
$\mathcal{M}_{spatial}=\mathcal{A}_{det,j}/\mathcal{A}_{coh}\gg1$, and temporal modes $\mathcal{M}_{t}=\mathcal{T}_{det}/\mathcal{T}_{coh}\gg1$
($\mathcal{T}_{det}$  detection time, $\mathcal{T}_{coh}$ the coherence time of PDC). The statistics of the detected light is, in principle, multi-thermal with mean value $\langle
N_{j}\rangle=\mathcal{M}_{tot} \eta_{j} \mu$ and
variance $\langle (\delta N_{j})^2\rangle =\mathcal{M}_{tot} \eta_{j} \mu ( 1 + \eta_{j} \mu) $, where
$\mathcal{M}_{tot}=\mathcal{M}_{t}\mathcal{M}_{spatial}$ and  $\eta_{s}$ ,
$\eta_{i}$ the detection efficiency of signal and idler beam respectively \cite{Brida2010}.

More practically, here we consider two CCD-sensor's regions $\mathcal{B}_i$ and $\mathcal{B}_s$ containing a finite amount of pixels corresponding approximately to the regions $\mathcal{A}_ {det,s}$ and $\mathcal{A}_ {det,i}$, respectively. $\mathcal{B}_i$ and $\mathcal{B}_s$ collect a certain number of
correlated modes $\mathcal{M}_c$,  and uncorrelated modes $\mathcal{M}_u$. Moreover, since some of the modes close to the borders of the detection areas are only partially collected, we introduce their number $\mathcal{M}_b$  and their mean collection efficiency $\beta$ (see Fig. 3).

The statistics of the photon number distribution obeys the following equations:
\begin{eqnarray}
\langle N_j \rangle &=& (\mathcal{M}_c + \mathcal{M}_u + \mathcal{M}_b \beta) \eta_j \mu , ~~~   ( j=1,2 ) \nonumber \\
%\langle (\delta N_j)^2 \rangle &=& (\mathcal{M}_c + \mathcal{M}_u + \mathcal{M}_b \beta) \eta_j \mu  + (\mathcal{M}_c + \mathcal{M}_u + \mathcal{M}_b \beta^2) (\eta_j \mu)^2, \\
\langle (\delta N_j)^2 \rangle &=& (\mathcal{M}_c + \mathcal{M}_u ) \eta_j \mu(1+ \eta_j \mu)  + \mathcal{M}_b   \beta  \eta_j \mu(1+ \beta  \eta_j \mu), \nonumber \\
\langle \delta N_s \delta N_i \rangle &=& (\mathcal{M}_c +  \mathcal{M}_b \beta^2) \eta_s \eta_i \mu (1+\mu).
\label{mean}
\end{eqnarray}
Here $\eta_j$ is the detection efficiency of the beam $j$ at 812nm, including  both the transmission efficiency and the CCD camera quantum efficiency. Moreover, we introduce the quantity $\alpha = \langle N_i \rangle / \langle N_s \rangle = \eta_i/ \eta_s$. $\alpha$ can be easily estimated and allows balancing, a posteriori, the photon counts of the two channels in the evaluation of the noise reduction factor $\sigma_{\alpha} \equiv \langle [\delta(N_i- \alpha N_s)]^2 \rangle / \langle N_i + \alpha N_s \rangle$  \cite{Brida2010}. By using Eq.s (\ref{mean})  it turns out that the balanced noise reduction factor $\sigma_\alpha$  can be written as:
\begin{equation}
\sigma_\alpha=  \frac{(1+ \alpha)}{2} - \eta_1 A \label{EQsigmaalfa}
\end{equation}
where
$A= (\mathcal{M}_c + \mathcal{M}_b \beta^2  - \mathcal{M}_u \mu) / ( \mathcal{M}_c + \mathcal{M}_u + \mathcal{M}_b \beta)$

By trivial geometrical arguments, for a square detection areas of size $L \times L$, with $L \gg r$, $L \gg d$,  the number of modes of the different types result to be estimated as $\mathcal{M}_u = (2 L d)(\pi r^2)^{-1}$,   $\mathcal{M}_b = (2 L )/r$,   $\mathcal{M}_c = (L^2 - 2 L d )(\pi r^2)^{-1}$,
where $d$ is the deviation of the centering of the pixel grid  with respect to the physical center of symmetry and $r$ is the radius of the coherence area at the detection plane.  The average collection efficiency is assumed to be $\beta=1/2$.
We observe that ideally $A$ is equal to one since both $\mathcal{M}_u$ and $\mathcal{M}_b$ are zero. In practice this condition cannot be achieved but
$A$ can be reliably estimated by careful measurements of the geometric parameters $L$, $d$, and $r$. In general, also the mean photon number per mode $\mu$ should be carefully evaluated to estimate $A$. However, being in our case $\mathcal{\mu}<10^{-7}$, it does not provide any significant contribution when $A$ is estimated.

Thus, from Eq. (\ref{EQsigmaalfa}) it is possible to extract the quantum efficiency with the needed accuracy, by evaluating experimentally $\sigma_\alpha$ , $\alpha$ , $A(L,r,d)$, and by properly selecting the correlated detector regions $\mathcal{B}_s$ and $\mathcal{B}_i$.

\begin{figure}[tbp]
\begin{center}
\includegraphics[angle=0, width=7cm]{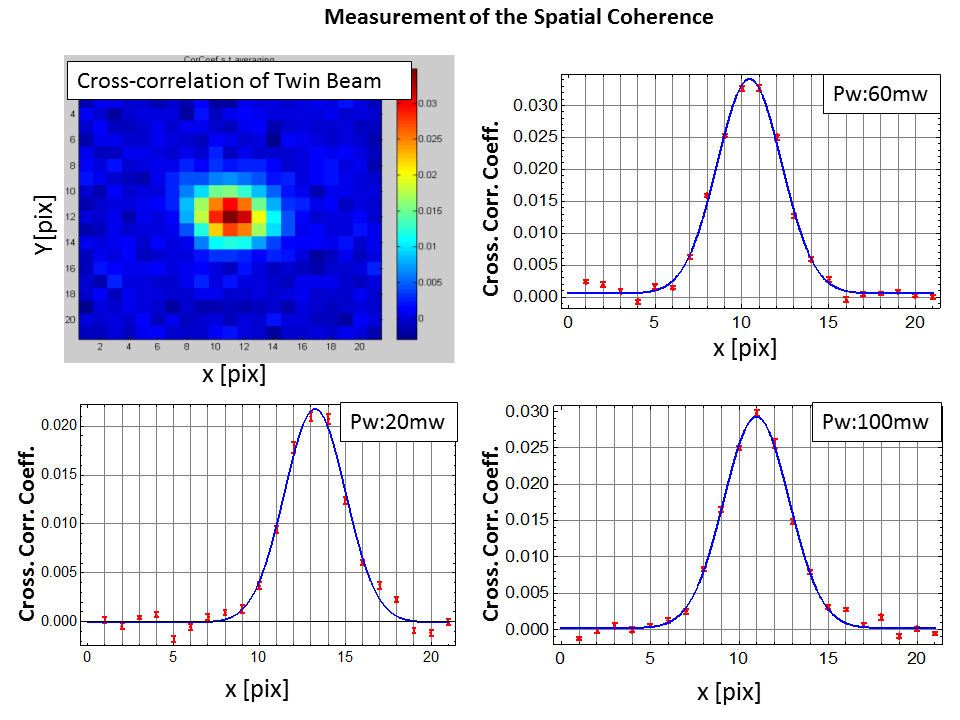}
\caption{Experimental evaluation of the  spatial cross- correlation coefficient  and the coherence radius. Top-left frame shows the 2-dimensional function  for $\mathbf{\xi}=(x,y)$, while the other graphs present horizontal sections.  "$P_w$" refers to different nominal power of the laser pump. \label{SpatialCoherence}}
\end{center}
\end{figure}

\emph{Coherence radius estimation.} We estimate the coherence radius as the FWHM of the spatial cross-correlation coefficient estimated experimentally by the formula
\begin{equation}
c(\mathbf{\xi})=\sum_\mathbf{x} \frac{\langle \delta N_i (\mathbf{x}) \delta N_s (-\mathbf{x}+\mathbf{\xi}) \rangle }{\sqrt{\langle [\delta N_i (\mathbf{x})]^2  \rangle   \langle [\delta N_s (-\mathbf{x}+\mathbf{\xi}))]^2  \rangle}}   \label{C(xi)}
\end{equation}
where $\mathbf{x}$ is the vector position of a pixel belonging to the large region considered. $\mathbf{\xi}$  represents the 2-dimensional shift of the second (correlated) region (one pixel steps, as shown e.g. in Fig. 3). For this measurement we use the highest resolution provided by the CCD camera, i.e. we take the images without pixel binning. The physical pixel size is 20 $\mu$m.  The correlation coefficient in function of the shift $\mathbf{\xi}=(x,y)$ is reported in Fig. \ref{SpatialCoherence}. The measure has been performed for different powers of the pump, in order to verify the reliability of the measurement. For all the power we obtain $r_0=43 (3)$  $\mu$m.

\begin{figure}[tbp]
\begin{center}\label{Modes}
\includegraphics[angle=0, width=7 cm]{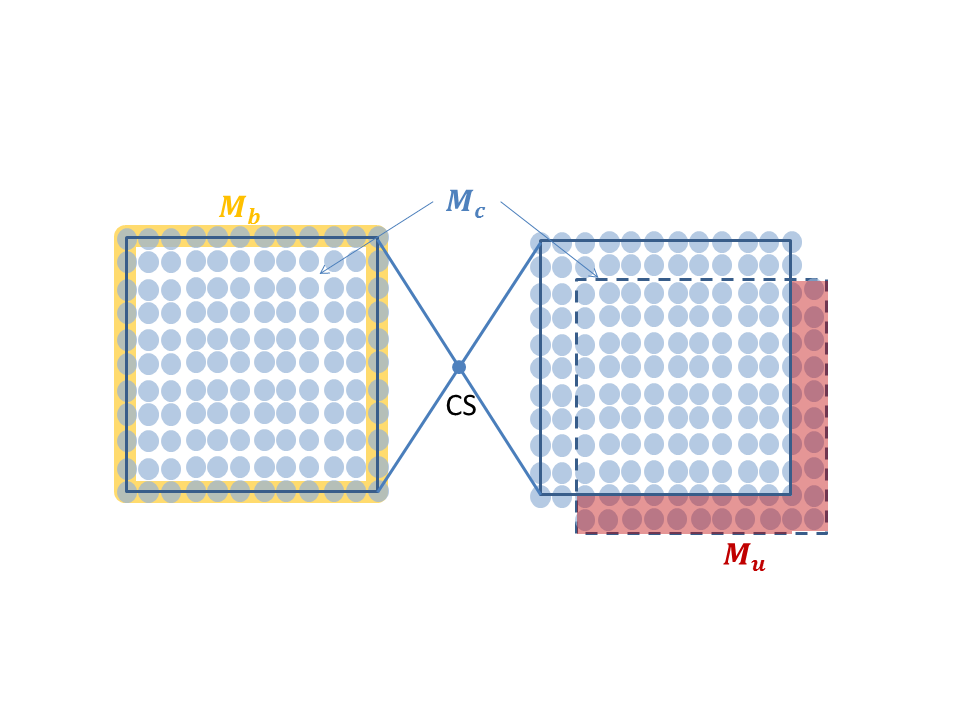}
\caption{A graphical visualization of the correlated modes $\mathcal{M}_c$, the uncorrelated $\mathcal{M}_u$ and the border modes $\mathcal{M}_b$ collected by CCD sensor's areas $\mathcal{B}_s$ and $\mathcal{B}_i$ when misaligned with respect to the Center-of-Simmetry (CS) of the correlated modes (blue circles)}
\end{center}
\end{figure}

\begin{figure}[tbp]
\begin{center}\label{Centering}
\includegraphics[angle=0, width=6cm]{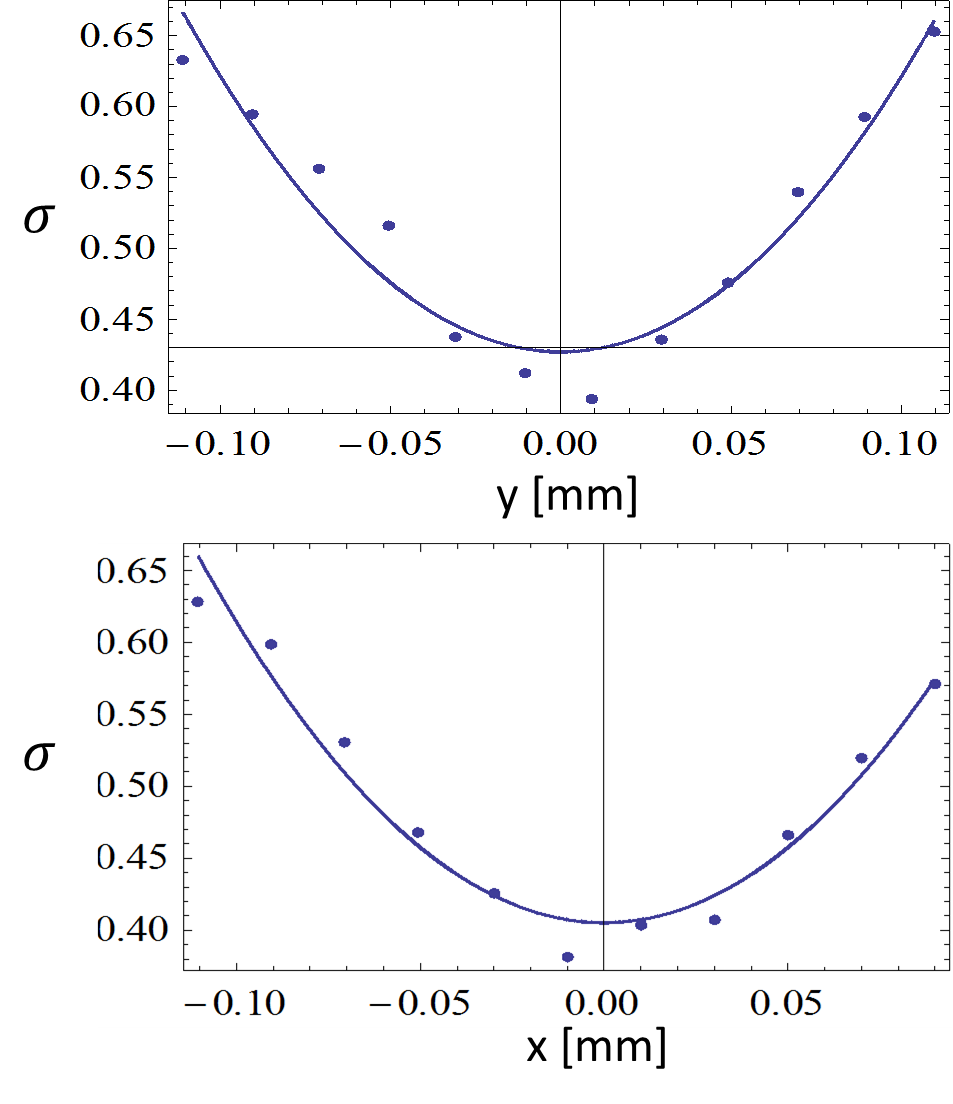}
\caption{Noise reduction factor of a the pairs of pixels versus the position of the CCD. \label{NRFxyCCD}}
\end{center}
\end{figure}

\emph{Alignment of the CCD sensor.} The aim is to reduce as much as possible the uncertainty on the optimal alignment, represented by $d=0$. After a first rough identification of two correlated regions in the detection plane  (indicated by blue squares in Fig. \ref{setup}) a fine procedure is used to centering the greed of pixels with respect to the physical center of symmetry of the degenerate emission. It consists in moving the CCD with sub-pixel and sub coherence resolution. For each step, the average noise reduction factor $\sigma \equiv \langle [\delta(N_i-N_s)]^2 \rangle / \langle N_i + N_s \rangle$ \cite{Brida2010}, between pairs of symmetric super-pixels, is evaluated. A double scan first in $x$ (horizontal axis) and then in $y$ (vertical axis) allows producing the two data set reported in Fig. \ref{NRFxyCCD}. Moving from the optimal position determines an increasing of $\sigma$ due to the appearing of uncorrelated photons in symmetric pixels pairs of the two areas indicated in blue in Fig. \ref{setup}: each pixel of a pair collects photons that are no more conjugated with the ones collected by its symmetric brother \cite{footnote1}. By a parabolic fit, we get the position of the minimum with an accuracy  of $\delta d = 2.7$ $\mu$m both in $x$ and $y$ directions.

\emph{Efficiency estimation.}
Integrating the number of photons $N_i$ and $N_s$ detected in optimally correlated areas $\mathcal{B}_s$ and $\mathcal{B}_i$ of each image, the balancing factor $\alpha$ and the noise reduction factor $\sigma_\alpha$ are then evaluated by averaging over 15000 images. The largest area available in our configuration corresponding to $L_0=2630$ $\mu$m (with negligible uncertainty). The coefficient $A$ in Eq. (\ref{EQsigmaalfa}) is $\mathcal{A}(L_0,d=0, r_0)= 0.9756 (0.0025)$, not far from the ideal value of 1. Fixing $L_0$ , measuring the parameter $\alpha= \langle N_i \rangle/ \langle N_s \rangle = 0.91867 (10^{-5})$, and  $\sigma_\alpha (L_0) = 0.253 (0.003)$ allows to estimating the quantum efficiency of channel $i$ by Eq. (\ref{EQsigmaalfa}), obtaining
\begin{equation}
\eta_i=0.724 (0.004) \label{eta1}
\end{equation}
From the value of $\alpha= \eta_i/ \eta_s$  it is possible to estimate $\eta_s=0.788 (0.004)$.

\begin{figure}[tbp]
\begin{center}
\includegraphics[angle=0, width=7cm]{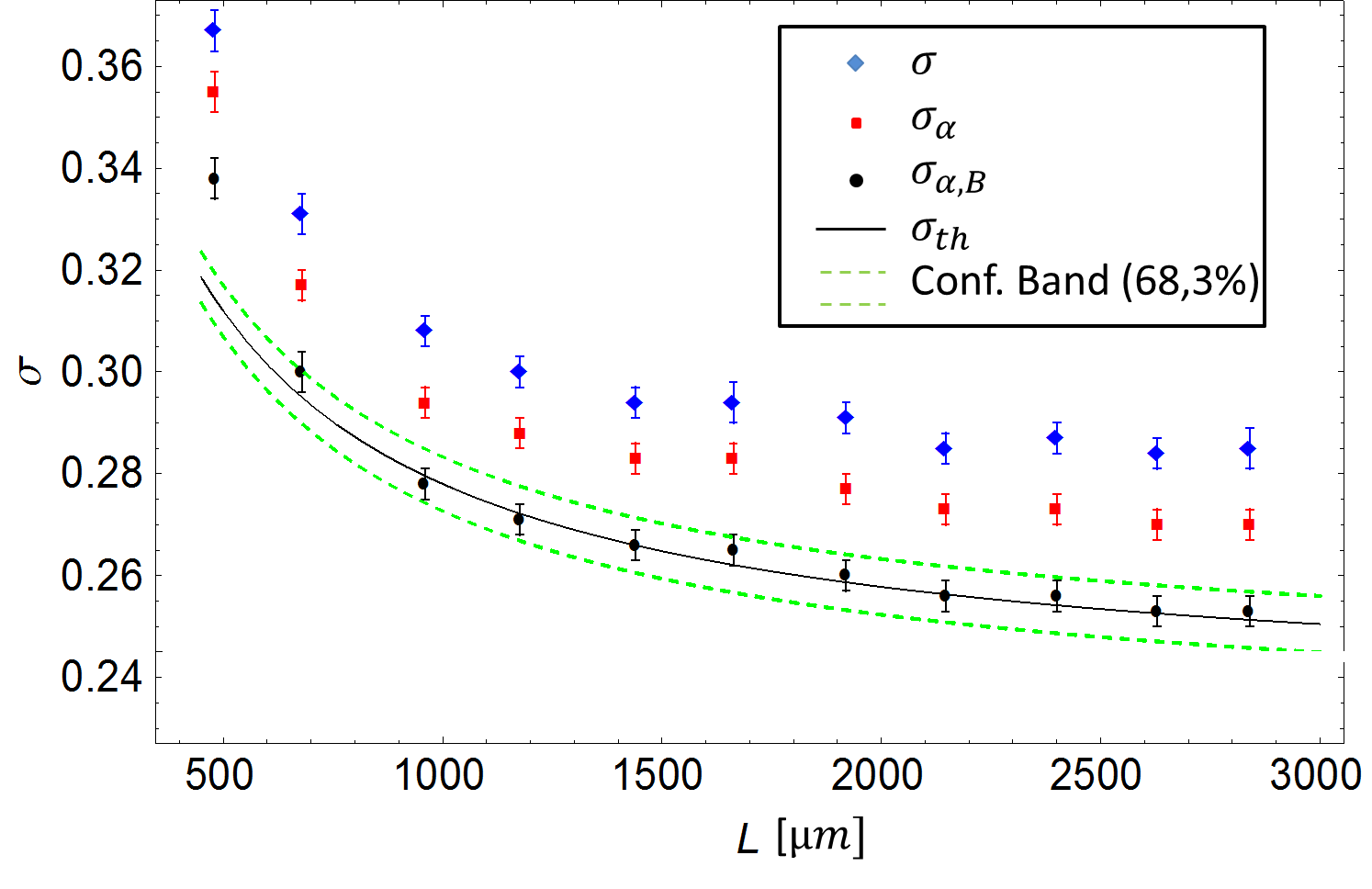}
\caption{ Noise reduction factor as a function of the linear size of the detection areas $L$. Here, uncertainty bars represent only the statistical fluctuations. See text for further details. \label{sigmas} }
\end{center}
\end{figure}

Even if this is already a remarkable result, it is possible to further reduce the uncertainty by a more sophisticated data analysis. In particular we measure $\sigma_\alpha$ as a function of the size of the detection areas $L$, from 2$\times$1 to 5$\times$6 super-pixels.  The results are shown in Fig. \ref{sigmas}, where we plot the raw noise reduction factor $\sigma$, the balanced one $\sigma_\alpha$, and the balanced and background (electronic noise, stray-light)  corrected one $\sigma_{\alpha,B}$. At the same time, each size of the detection area gives a different value of the corrective coefficient $A(r,d,L)$, with $L$ ranging from $L_1=679$ $\mu$m to $L_{10}=2630$ $\mu$m.

Fig. \ref{sigmas} shows also the theoretical curve of the noise reduction factor calculated by introducing into Eq. (\ref{EQsigmaalfa})  the value of the efficiency of Eq. (\ref{eta1}), and its confidence interval. In general the data are distributed in agreement with the theory, validating of our theoretical model.

\begin{figure}[tbp]
\begin{center}\label{SigmaVsA}
\includegraphics[angle=0, width=7cm]{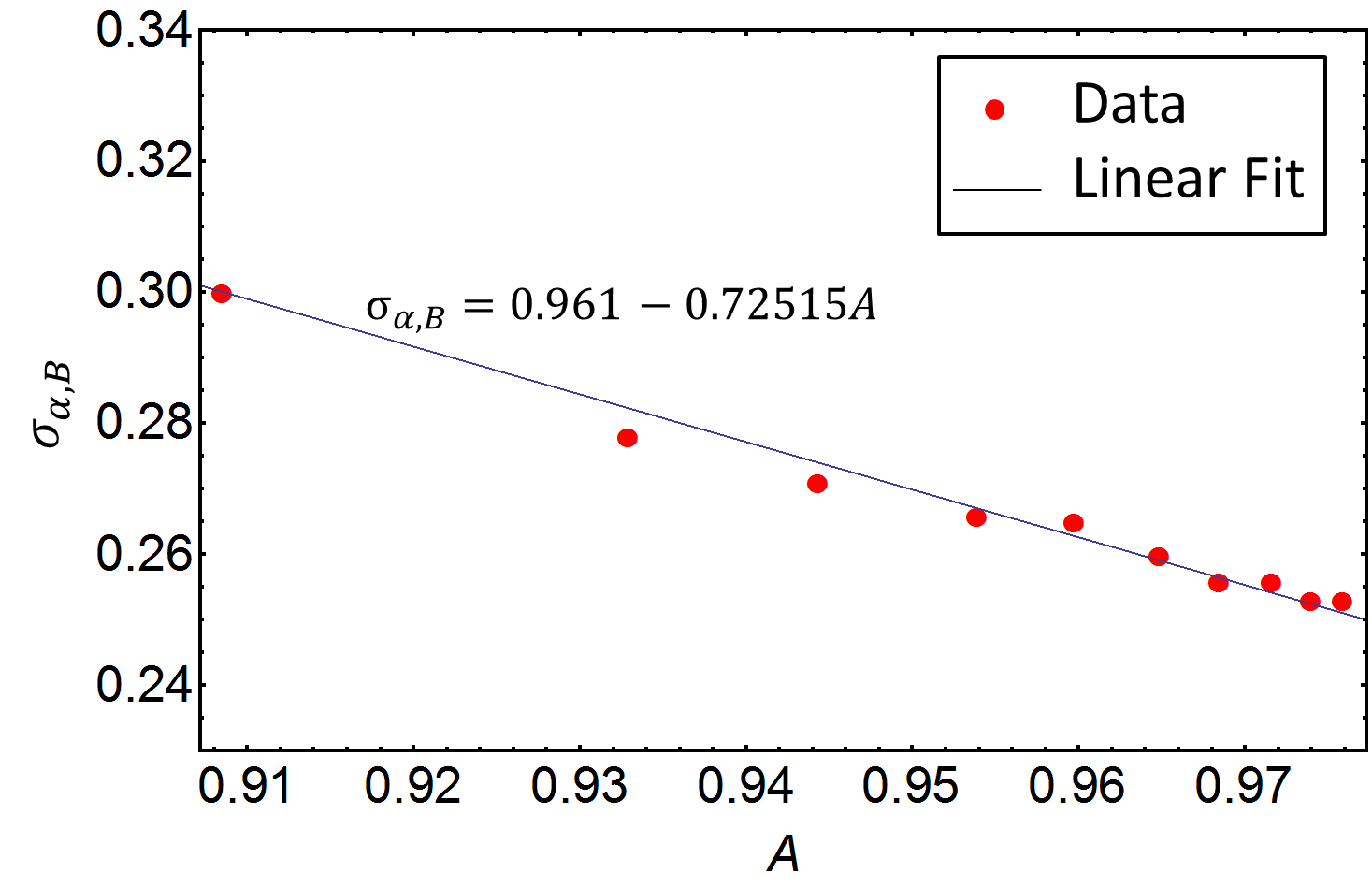}
\caption{Dependence of the balanced noise reduction factor from the geometrical correction coefficient A and the corresponding equation of the linear fit.\label{SigmaVsA} }
\end{center}
\end{figure}

In Fig. \ref{SigmaVsA} we present the dependence of the noise reduction factor $\sigma_{\alpha, B}$ on the coefficient $\mathcal{A}$.  Again, as predicted by our  model summarized in Eq. (\ref{EQsigmaalfa}), the relation is linear. The fit of the data, with the quantum efficiency $\eta_i$ as the only free parameter,  provides in fact a correlation coefficient of  0.9999.

Each point of the plot representing a pair  $A(L), \sigma_{\alpha,B}(L)$ delivers a certain value of quantum efficiency by Eq. (\ref{EQsigmaalfa}). The final value of the quantum efficiency is taken as the arithmetic average $\overline{\eta}$ of the ten values. Note that the ten measurement of $\sigma_{\alpha,B}$ for different size $L$ are not independent since they are obtained from the same set of images, by growing the size of the areas around the same central point, i.e. the number of photons detected in a small area, and the photon counts in a larger one are somehow correlated. This is accounted for in the evaluation of the uncertainty of $\overline{\eta}$, starting from Eq. (\ref{EQsigmaalfa}) and evaluating the corresponding variance-covariance matrix, thus obtaining \cite{footnote2}:
\begin{equation}
\overline{\eta_i}= 0.725 (0.002).
\end{equation}
where the relative uncertainty of 0.3\% is reduced whit respect to the single point measurement reported in Eq. (\ref{eta1}), without reaching the enhanced factor $1/\sqrt{10}$ expected for the case of independent measurements.

\emph{Conclusions.}
We heve demonstrated an effective and practical method for the absolute calibration of a CCD camera with high accuracy ($0.3\%$ relative uncertainty), exploiting squeezed vacuum and an analog detection regime of thousands of photons. Being absolute, the method does not require comparison with standards, whereas it relies on fundamental physical laws. Any known uncertainty contribution have been carefully evaluated and corrected for, making the present methods competitive and at the same time complementary with respect other methods of traditional radiometry and quantum radiometry (e.g. Klyshko's technique) in the same regime. Moreover, modern electro-multiplied CCD are valuable devices capable of working with high efficiency from the single photon to the analog macroscopic regime, that have become largely used in quantum information \cite{BlanchetPRL2008,EdgarNC2012,FicklerSR2013} and quantum imaging \cite{AspdenNJP2013,qm2, BridaNP2010}. This makes our achievement a useful tools for these emerging technologies.

{\bf Acknowledgements}

has received funding from the European Union BRISQ project, JRP EXL02 - SIQUTE project (on the basis of Decision No.
912/2009/EC), from Fondazione San Paolo, from MIUR (FIRB ``LiCHIS'' - RBFR10YQ3H and Progetto Premiale  ``Oltre i limiti classici di misura''), and from NATO grant EAP-SFPP98439.

%\tableofcontents
%\section*{References}


\begin{thebibliography}{10}
\bibitem{footnote1} Note that the minimum value of the noise reduction factor $\sigma$ in Fig. \ref{NRFxyCCD} is not yet limited only by quantum efficiency. Alignment procedure has minimize the uncorrelated modes, however the effect of the border modes $\mathcal{M}_{b}$ is still large for a detection area of one superpixel $l_{pix}=$480 $\mu$m.
\bibitem{footnote2}The variance associated to the measurement of $\mathcal{A}$ ranges from 6.7 $10^{-5}$ to 5.5 $10^{-6}$, due to uncertainty on the determination of the center of symmetry and of the radius of the speckle; the variance and the covariance on $\alpha$ are both of the order of ($\sim 10^{-9}$). A great contribution to the covariance matrix cames from $\sigma$, with both variances and covariances of $\sim 10^{-5}$ for the smallest area scaling up to $\sim 10^{-4}$ for the largest one.
\end{thebibliography}
\end{document}